\documentclass[english]{IEEEtran}
\usepackage[T1]{fontenc}
\usepackage[latin9]{inputenc}
\usepackage{color}
\usepackage{babel}
\usepackage{float}
\usepackage{amsmath}
\usepackage{amssymb}
\usepackage{graphicx}
\usepackage{microtype}
\usepackage[unicode=true,pdfusetitle,
 bookmarks=true,bookmarksnumbered=false,bookmarksopen=false,
 breaklinks=false,pdfborder={0 0 1},backref=false,colorlinks=false]
 {hyperref}

\makeatletter

\floatstyle{ruled}
\newfloat{algorithm}{tbp}{loa}
\providecommand{\algorithmname}{Algorithm}
\floatname{algorithm}{\protect\algorithmname}

\usepackage{flushend}

\@ifundefined{showcaptionsetup}{}{%
 \PassOptionsToPackage{caption=false}{subfig}}
\usepackage{subfig}
\makeatother

\begin{document}
\title{Reconfigurable Intelligent Surface Optimization for Uplink Sparse
Code Multiple Access}
\author{Ibrahim Al-Nahhal, \textit{Member, IEEE},\textit{ }Octavia A. Dobre,
\textit{Fellow, IEEE}, and Ertugrul Basar, \textit{Senior Member,
IEEE, }Telex M. N. Ngatched, \textit{Senior Member, IEEE,} and Salama
Ikki, \textit{Senior} \textit{Member, IEEE}\thanks{This work was supported by the Natural Sciences and Engineering Research
Council of Canada (NSERC), through its Discovery program. The work
of E. Basar was supported by TUBITAK under Grant 120E401.}\thanks{O. A. Dobre, I. Al-Nahhal and T. M. N. Ngatched are with the Faculty
of Engineering and Applied Science, Memorial University, St. John\textquoteright s,
NL, Canada, (e-mail: \{odobre, ioalnahhal\}@mun.ca; tngatched@grenfell.mun.ca).}\thanks{E. Basar is with the CoreLab, Department of Electrical and Electronics
Engineering, Ko\c{c} University, Istanbul, Turkey (e-mail: ebasar@ku.edu.tr).}\thanks{S. Ikki is with the Department of Electrical Engineering, Lakehead
University, Thunder Bay, ON, Canada (e-mail: sikki@lakeheadu.ca).}\\
\vspace{-5.5mm}
}
\maketitle
\begin{abstract}
The reconfigurable intelligent surface (RIS)-assisted sparse code
multiple access (RIS-SCMA) is an attractive scheme for future wireless
networks. In this letter, for the first time, the RIS phase shifts
of the uplink RIS-SCMA system are optimized based on the alternate
optimization (AO) technique to improve the received signal-to-noise
ratio (SNR) for a discrete set of RIS phase shifts. The system model
of the uplink RIS-SCMA is formulated to utilize the AO algorithm.
For further reduction in the computational complexity, a low-complexity
AO (LC-AO) algorithm is proposed. The complexity analysis of the two
proposed algorithms is performed. Monte Carlo simulations and complexity
analysis show that the proposed algorithms significantly improve the
received SNR compared to the non-optimized RIS-SCMA scenario. The
LC-AO provides the same received SNR as the AO algorithm, with a significant
reduction in complexity. Moreover, the deployment of RISs for the
uplink RIS-SCMA is investigated.
\end{abstract}


\vspace{-5mm}

\section{Introduction}

\def\figurename{Fig.}
\def\tablename{TABLE}

\IEEEPARstart{R}{econfigurable} intelligent surface (RIS)-empowered
communication has recently become a promising candidate technology
for future wireless networks. An RIS consists of low-cost passive
elements that reflect the incident signals after adjusting their phases
and/or amplitudes \cite{RIS_Ertugrul 2019,RIS_2020}. By intelligent
configuration of the incident signals' phases and/or amplitudes, the
constructive and destructive interference between the reflected signals
can be manipulated. Thus, the transmission environment of the wireless
medium and quality-of-service can be improved without the need for
coding \cite{RIS_3}.

Sparse code multiple access (SCMA) is a code-domain non-orthogonal
multiple access (NOMA) approach that has received considerable attention
from the research community in the past few years \cite{Mohammadkarimi_Octavia}-\cite{SM-SCMA_Complexity}.
SCMA provides a spectrally efficient transmission by assigning unique
sparse codes to users who share the wireless medium \cite{Nikopour_SCMA_2013,codebook_design_2014}.
The codes' sparsity property enables the use of the iterative message
passing algorithm (MPA) at the receiver, to provide a near optimum
decoding performance with an implementable decoding complexity \cite{MPA_2015}.

Power-domain NOMA is another variant of NOMA in which users' signals
are superposed with different power levels \cite{PD-NOMA_new}. Power-domain
NOMA assisted by the RIS technology has been explored in \cite{NOMA_RIS_1}-\cite{NOMA_RIS_3}.
In \cite{NOMA_RIS_1} and \cite{MIMO-NOMA-RIS}, the downlink system
performance for the RIS-NOMA scheme is improved for single and multiple
antenna systems, respectively. The authors in \cite{NOMA_RIS_2} and
\cite{NOMA_RIS_2_1} utilize the RIS to maximize the sum-rate for
the downlink millimeter-wave NOMA and uplink NOMA, respectively. The
sum coverage range maximization is performed in \cite{NOMA_RIS_3}
for a downlink RIS-NOMA system. On the other hand, SCMA and RIS have
only been recently explored in \cite{SCMA-RIS_Ibrahim,SCMA-IRS}.
The first investigation in \cite{SCMA-RIS_Ibrahim} utilizes the RIS
technology for reducing the decoding complexity of the conventional
uplink SCMA system without optimizing the RIS phase shifts. In contrast,
the authors in \cite{SCMA-IRS} provide the error rate and sum-rate
performance analysis for the uplink RIS-SCMA system. To the best of
the authors' knowledge, no published work has investigated the optimization
of the RIS phases in detail for RIS-SCMA.

This letter provides two algorithms to optimize the RIS discrete phase
shifts of the uplink RIS-SCMA system, for the first time. These algorithms
are based on alternate optimization (AO), and are referred to as AO
and low-complexity AO (LC-AO). It is worth noting that employing the
AO algorithm for the uplink RIS-SCMA requires mathematical manipulation
to formulate the system model. The LC-AO algorithm is proposed to
provide the same performance as the AO algorithm, but with a significant
reduction in the computational complexity. The proposed algorithms
considerably improve  the received signal-to-noise ratio (SNR) compared
to the non-optimized (i.e., blind) scenario by iteratively optimizing
the discrete RIS phase shifts. Monte Carlo simulations and complexity
analysis are provided to assess the proposed algorithms.

The remainder of this letter is organized as follows: Section \ref{sec:Uplink-SCMA-RIS-System}
presents the system model of the uplink RIS-SCMA system. The formulation
of the optimization problem and the proposed algorithms are introduced
in Section \ref{sec:Proposed-RIS-phase}, while their complexity is
analyzed in Section \ref{sec:Complexity-Analysis}. Finally, simulation
results and conclusions are presented in Sections \ref{sec:Simulation-Results}
and \ref{sec:Conclusion}, respectively.

\vspace{-2mm}

\section{\label{sec:Uplink-SCMA-RIS-System}Uplink RIS-SCMA System Model}

\begin{figure}
\begin{centering}
\includegraphics[scale=0.31]{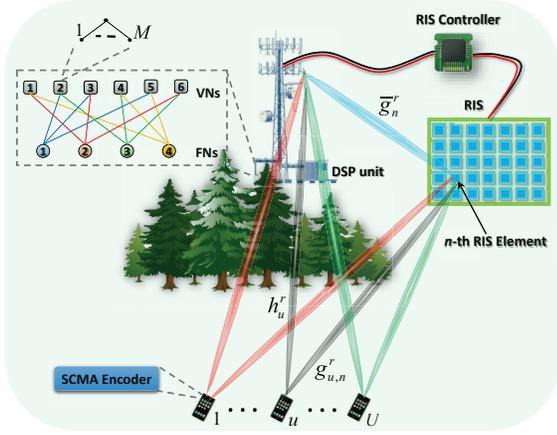}
\par\end{centering}
\caption{\textcolor{blue}{\label{fig: SCMA_RIS_BlockDiagram}}Uplink RIS-SCMA
system.}

\vspace{-5mm}
\end{figure}

Fig. \ref{fig: SCMA_RIS_BlockDiagram} illustrates an uplink RIS-SCMA
system that consists of $U$ uplink single-antenna users that encode
their data using unique sparse codebooks, $\mathbf{C}_{u}\in\mathbb{C}^{R\times M}$,
$u=1,\ldots,U$, where $M$ and $R$ (for code-domain NOMA, $U>R$)
are the size of the user's codebook and orthogonal resource elements
(OREs), respectively. The user's codebook, $\mathbf{c}_{u,m}\in\mathbb{C}^{R\times1}$,
$m=1,\ldots,M$, contains $d_{v}$ non-zero codeword elements. It
is worth noting that the sparse codebooks for all users are designed
to share the same $R$ OREs in which the number of non-zero shared
codeword elements for each ORE, $d_{f}$, is fixed. It is assumed
that each single-antenna user delivers its data to a single-antenna
base station (BS) through a line-of-sight (LoS) path and through an
$N$ passive reflecting RIS elements.

The ORE for SCMA can be a frequency-unit or a time-unit depending
on the nature of the problem. In this paper, a time-unit ORE is considered
for the RIS-SCMA systems. For each ORE, the received signal, $y^{r}$,
at the receiver-side is given by

\vspace{-5mm}

\begin{equation}
y^{r}=\sum_{u\in\varLambda_{r}}\left(\sum_{n=1}^{N}\left(g_{u,n}^{r}e^{-\jmath\phi_{n}}\bar{g}_{n}^{r}+h_{u}^{r}\right)\right)c_{u,m}^{r}+w^{r},\label{eq: y_r}
\end{equation}

\noindent where $h_{u}^{r}$, $\bar{g}_{n}^{r}$ and $g_{u,n}^{r}$
are the channel coefficients of the $u$-th user to BS (i.e., the
LoS component), $n$-th RIS element to BS, and $u$-th user to $n$-th
RIS element, respectively, at the $r$-th ORE. $c_{u,m}^{r}$ is the
$u$-th user's codeword at the $r$-th ORE. The additive white Gaussian
noise at the $r$-th ORE is denoted as $w^{r}\sim\mathcal{N}(0,\sigma_{r}^{2})$
with zero-mean and variance of $\sigma_{r}^{2}$; for simplicity,
it is assumed that the noise variance for all $R$ OREs is the same.
$\varLambda_{r}$ denotes the users' indices that are interfering
over the $r$-th ORE. Here, $\phi_{n}$ represents the $n$-th discrete
phase shift of the RIS reflecting at the $r$-th ORE, and its value
is taken from a finite set, $\mathcal{L}\in[-\pi,\pi)$, as

\vspace{-5mm}

\begin{equation}
\phi_{n}\in\mathcal{L}=\left\{ -\pi,\,-\pi+\Delta,\,\ldots,\,-\pi+(2^{b}-1)\Delta\right\} \,\,\,\,\,\forall n,\label{eq: Discrete Phases}
\end{equation}

\noindent where $b$ represents the number of quantization bits for
each RIS phase shift, and $\Delta=2\pi/2^{b}$ is the step size between
 two successive discrete values of the RIS phase shifts. At the receiver,
the MPA \cite{SCMA-RIS_Ibrahim} is employed to decode the users'
messages by iteratively updating the messages between function nodes
(FN) and variable nodes (VN), as illustrated in Fig. \ref{fig: SCMA_RIS_BlockDiagram}.

The received SNR at the $r$-th ORE, $\Gamma^{r}$, is

\vspace{-5mm}

\begin{equation}
\Gamma^{r}=\frac{E}{\sigma_{r}^{2}}\left|\sum_{u\in\varLambda_{r}}\left(\sum_{n=1}^{N}\left(g_{u,n}^{r}e^{-\jmath\phi_{n}}\bar{g}_{n}^{r}+h_{u}^{r}\right)\right)\right|^{2},\label{eq: Rx SNR}
\end{equation}

\noindent where $E$ is the average transmitted user energy. Here,
the optimization of the RIS phase shifts aims to maximize the received
SNR at the $r$-th ORE, as

\vspace{-4mm}

\begin{equation}
\text{(P1):}\,\,\,\,\,\,\,\underset{\begin{array}{c}
\phi_{n}\end{array}}{\text{max}}\left|\sum_{u\in\varLambda_{r}}\left(\sum_{n=1}^{N}\left(g_{u,n}^{r}e^{-\jmath\phi_{n}}\bar{g}_{n}^{r}+h_{u}^{r}\right)\right)\right|^{2}\label{eq: P1}
\end{equation}

\vspace{-2mm}

\begin{equation}
\text{s.t.}\,\,\,\,\,\,\phi_{n}\in\mathcal{L},\,\,\,\,\forall n,\,\,\forall r.\hspace{20mm}\label{eq: P1_C1}
\end{equation}

\noindent The optimization problem in (P1) is non-convex due to the
objective function itself and the finite values of the phase shifts.
There is no unique method to relax a non-convex problem to obtain
an optimal solution efficiently. The optimal solution can be achieved
by utilizing an exhaustive search for all combinations of the discrete
RIS phase shifts. However, the computational complexity of this approach
is practically prohibitive. It is worth noting that optimizing the
received SNR can be considered for the RIS-SCMA system due to the
shaping gain of the users' sparse codebooks.

\vspace{-2mm}

\section{\label{sec:Proposed-RIS-phase}Proposed RIS phase shifts Optimization}

As mentioned before, finding the optimal set of RIS phase shifts requires
trying $\sim(2^{bN}R)$ possible combinations, which is extremely
costly and infeasible even for a small-scale problem. A sub-optimal
solution with acceptable computational complexity can be obtained
by utilizing the AO concept. It should be emphasized that the AO algorithm
is adopted in \cite{AO-Beamforming} for a different wireless communication
scheme, here, a reformulation of (P1) is needed to apply the AO algorithm.

\vspace{-3mm}

\subsection{Proposed AO Algorithm}

In the proposed AO algorithm, the solution of (P1) in (\ref{eq: P1})
can be obtained by iteratively optimizing the RIS phase shifts one
at a time at each ORE. Consider $\bar{\boldsymbol{g}}_{n}^{r}\in\mathbb{C}^{1\times N}=[\bar{g}_{1}^{r}\ldots\bar{g}_{n}^{r}\ldots\bar{g}_{N}^{r}]$,
$\Phi^{r}\in\mathbb{C}^{N\times N}=\text{diag}([e^{-\jmath\phi_{1}}\ldots e^{-\jmath\phi_{n}}\ldots e^{-\jmath\phi_{N}}])$
with $\text{diag}(\cdot)$ reshaping the corresponding vector to a
diagonal square matrix, $\boldsymbol{G}^{r}\in\mathbb{C}^{N\times d_{f}}=[\boldsymbol{g}_{1}^{r}\ldots\boldsymbol{g}_{u}^{r}\ldots\boldsymbol{g}_{d_{f}}^{r}]$
with $\boldsymbol{g}_{u}^{r}\in\mathbb{C}^{N\times1}=[g_{u,1}^{r}\ldots g_{u,n}^{r}\ldots g_{u,N}^{r}]^{\dagger}$
$\forall u\in\varLambda_{r}$, $\boldsymbol{h}^{r}\in\mathbb{C}^{1\times d_{f}}=[h_{1}^{r}\ldots h_{u}^{r}\ldots h_{d_{f}}^{r}]$
$\forall u\in\varLambda_{r}$, and $\boldsymbol{c}^{r}\in\mathbb{C}^{d_{f}\times1}=[c_{1,m}^{r}\ldots c_{u,m}^{r}\ldots c_{d_{f},m}^{r}]^{\dagger}$
$\forall u\in\varLambda_{r}$, with $\dagger$ denoting transpose.
Therefore, $y^{r}$ in (\ref{eq: y_r}) and $\Gamma^{r}$ in (\ref{eq: Rx SNR})
can respectively be written as

\vspace{-3mm}

\begin{equation}
y^{r}=\left(\bar{\boldsymbol{g}}_{n}^{r}\Phi^{r}\boldsymbol{G}^{r}+\boldsymbol{h}^{r}\right)\boldsymbol{c}^{r}+w^{r},\label{eq: y_r norm}
\end{equation}

\noindent and

\vspace{-3mm}

\begin{equation}
\Gamma^{r}=\frac{E}{\sigma_{r}^{2}}\left\Vert \bar{\boldsymbol{g}}_{n}^{r}\Phi^{r}\boldsymbol{G}^{r}+\boldsymbol{h}^{r}\right\Vert ^{2}.\label{eq: Rx SNR-OA}
\end{equation}

The proposed AO algorithm employs an exhaustive search for only one
phase shift, while the remaining $N-1$ phase shifts are kept fixed.
An iterative procedure is required to improve the performance of the
proposed AO algorithm. The optimized phase shift of the $n$-th RIS
element for the $r$-th ORE, $\phi_{n}^{*}$, is

\vspace{-3mm}

\begin{equation}
\phi_{n}^{*}=\underset{\begin{array}{c}
\phi_{n,t}\in\mathcal{L}\end{array}}{\text{arg}\,\text{\,max}}\hspace{-2mm}\left\Vert \bar{\boldsymbol{g}}_{n}^{r}\Phi^{r}(\phi_{n,t})\boldsymbol{G}^{r}+\boldsymbol{h}^{r}\right\Vert ^{2},\,\,\,\,r=1,\ldots,R,\label{eq: theta_AO}
\end{equation}

\noindent where $\Phi^{r}(\phi_{n,t})=\text{diag}([e^{-j\phi_{1,t-1}}\ldots e^{-j\phi_{n,t}}\ldots e^{-j\phi_{N,t-1}}])$
is the phase shifts matrix with only varying $\phi_{n,t}$ at the
$t$-th iteration, while the rest of phase shifts are given by the
previous iteration (i.e., the $(t-1)$-th iteration). Finally, the
algorithm stops after $T$ iterations. Algorithm \ref{alg:The-proposed-AO}
summarizes the steps of the proposed AO algorithm.

\begin{algorithm}
\begin{itemize}
\item \textbf{Input:} $\bar{\boldsymbol{g}}_{n}^{r}$, $\boldsymbol{G}^{r}$,
$\boldsymbol{h}^{r}$, $\mathcal{L}$, and maximum \# of iteration,
$T$;
\item \textbf{Initiate:}{\small{} $\phi_{n,0}=0$} for \textbf{$n=1,\ldots,N$};
\end{itemize}
~~~~~1: \textbf{Set: $t\leftarrow1$};

~~~~~2: \textbf{for$\,\,\,r=1$ to $R$}

~~~~~3:\textbf{ ~~~while} \textbf{$\,\,\,t\leq T$, do}

~~~~~4:\textbf{ ~~~~~~~~~for$\,\,\,n=1$ to $N$}

~~~~~5:\textbf{ ~~~~~~~~~~~~Buffer: $\zeta_{max}=[\cdot]$};

~~~~~6:\textbf{ ~~~~~~~~~~~~for$\,\,\,l=1$ to $2^{b}$}

~~~~~7:\textbf{ ~~~~~~~~~~~~~~~~$\zeta_{max}(1,l)\leftarrow\left\Vert \bar{\boldsymbol{g}}_{n}^{r}\Phi^{r}(\phi_{n,t,l})\boldsymbol{G}^{r}+\boldsymbol{h}^{r}\right\Vert ^{2}$};

~~~~~8:\textbf{ ~~~~~~~~~~~~end for}

~~~~~9:\textbf{ ~~~~~~~~~~~~Find: $i_{max}=\text{arg\,\,max}\{\zeta_{max}\}$};

~~\hspace*{1mm}\hspace*{1mm}10: \textbf{~~~~~~~~~~~~Update:
$\phi_{n,t}\leftarrow\mathcal{L}(i_{max})$};

~~\hspace*{1mm}\hspace*{1mm}11:\textbf{ ~~~~~~~~~end
for}

~~\hspace*{1mm}\hspace*{1mm}12:\textbf{ ~~~~~~~~~Set:
$t\leftarrow t+1$};

~~\hspace*{1mm}\hspace*{1mm}13: \textbf{~~~end while}

~~\hspace*{1mm}\hspace*{1mm}14: \textbf{~~~Set: $\Phi^{r}\leftarrow\text{diag}([e^{-j\phi_{1,T}}\ldots e^{-j\phi_{N,T}}])$}

~~\hspace*{1mm}\hspace*{1mm}15: \textbf{end for}
\begin{itemize}
\item \textbf{Output}{\small{} }$\Phi^{r}$ for $r=1,\ldots,R$.
\end{itemize}
\caption{\label{alg:The-proposed-AO}The proposed AO algorithm pseudo-code.}
\end{algorithm}

\vspace{-3mm}

\subsection{Proposed LC-AO Algorithm}

As seen from line \#7 in Algorithm \ref{alg:The-proposed-AO}, the
AO algorithm calculates some terms that are not a function of $\phi_{n,t}$,
in each iteration. Thus, we can achieve the same performance as Algorithm
\ref{alg:The-proposed-AO} with a significant reduction in computational
complexity by eliminating these terms. In this subsection, the target
is to find and neglect these unnecessary terms.

To achieve this goal, consider $\bar{\boldsymbol{g}}_{n}^{r}\Phi^{r}\boldsymbol{G}^{r}=\boldsymbol{v}^{r}\text{diag}(\bar{\boldsymbol{g}}_{n}^{r})\boldsymbol{G}^{r}$,
where $\boldsymbol{v}^{r}\in\mathbb{C}^{1\times N}=[e^{-j\phi_{1}}\ldots e^{-j\phi_{N}}]$,
and $\boldsymbol{\Xi}^{r}\in\mathbb{C}^{N\times d_{f}}=\text{diag}(\bar{\boldsymbol{g}}_{n}^{r})\boldsymbol{G}^{r}$.
Thus, (\ref{eq: Rx SNR-OA}) can be written as

\vspace{-3mm}

\[
\Gamma^{r}=\frac{E}{\sigma_{r}^{2}}\left\Vert \boldsymbol{v}^{r}\boldsymbol{\Xi}^{r}+\boldsymbol{h}^{r}\right\Vert ^{2}\hspace{4.5cm}
\]

\vspace{-5mm}

\begin{equation}
=\frac{E}{\sigma_{r}^{2}}\left(\underset{\text{Term 1}}{\underbrace{\boldsymbol{v}^{r}\boldsymbol{D}^{r}(\boldsymbol{v}^{r})^{H}}}+\underset{\text{Term 2}}{\underbrace{2\Re\left\{ \boldsymbol{v}^{r}\bar{\boldsymbol{d}}^{r}\right\} }}+\left\Vert \boldsymbol{h}^{r}\right\Vert ^{2}\right),\,\,\label{eq: Rx SNR-LC-OA-1}
\end{equation}

\noindent where $\Re\{\cdot\}$ returns the real value of a complex
number, $\boldsymbol{D}^{r}\in\mathbb{C}^{N\times N}=\boldsymbol{\Xi}^{r}(\boldsymbol{\Xi}^{r})^{H}$,
and $\bar{\boldsymbol{d}}^{r}\in\mathbb{C}^{N\times1}=\boldsymbol{\Xi}^{r}(\boldsymbol{h}^{r})^{H}$.
Thus, the $(k,n)$-th element of $\boldsymbol{D}^{r}$ (i.e., $d_{k,n}^{r}$)
and the $k$-th element of $\bar{\boldsymbol{d}}^{r}$ (i.e., $\bar{d}_{k}^{r}$)
are respectively given as

\vspace{-1mm}

\begin{equation}
d_{k,n}^{r}=\sum_{i=1}^{d_{f}}\left(g_{k,i}^{r}\bar{g}_{k}^{r}\right)\left(g_{n,i}^{r}\bar{g}_{n}^{r}\right)^{*},\label{eq: d_k_n}
\end{equation}

\vspace{-1mm}

\noindent and

\vspace{-3mm}

\begin{equation}
\bar{d}_{k}^{r}=\sum_{i=1}^{d_{f}}\left(g_{k,i}^{r}\bar{g}_{k}^{r}\right)\left(h_{i}^{r}\right)^{*},\label{eq: d_k_hat}
\end{equation}

\vspace{-1mm}

\noindent where $(\cdot)^{*}$ denotes the conjugate of the complex
number.

Now, we need to find the terms in (\ref{eq: Rx SNR-LC-OA-1}) that
are not a function of $\phi_{n}$, to be neglected. Term 1 in (\ref{eq: Rx SNR-LC-OA-1})
can be written as

\vspace{-3mm}

\begin{equation}
\boldsymbol{v}^{r}\boldsymbol{D}^{r}(\boldsymbol{v}^{r})^{H}=A_{1}^{r}\left(\phi_{n}\right)+A_{1}^{r}\left(\Phi^{r}\backslash\phi_{n}\right),\label{eq: term 1}
\end{equation}

\noindent where

\vspace{-3mm}

\begin{equation}
A_{1}^{r}\left(\phi_{n}\right)=2\Re\left\{ e^{-j\phi_{n}}\sum_{k\neq n}^{N}e^{j\phi_{k}}\left(d_{k,n}^{r}\right)^{*}\right\} ,\label{eq: A_1_phi}
\end{equation}

\noindent and

\vspace{-3mm}

\begin{equation}
A_{1}^{r}\left(\Phi^{r}\backslash\phi_{n}\right)=\sum_{n=1}^{N}d_{n,n}^{r}+2\Re\left\{ \sum_{i\neq n}^{N}\sum_{j\neq n}^{N}e^{j(\phi_{i}-\phi_{j})}\left(d_{i,j}^{r}\right)^{*}\right\} .\label{eq: A_1_without_phi}
\end{equation}

\noindent Similarly, Term 2 in (\ref{eq: Rx SNR-LC-OA-1}) can be
written as

\vspace{-3mm}

\begin{equation}
2\Re\left\{ \boldsymbol{v}^{r}\bar{\boldsymbol{d}}^{r}\right\} =A_{2}^{r}\left(\phi_{n}\right)+A_{2}^{r}\left(\Phi^{r}\backslash\phi_{n}\right),\label{eq: term 2}
\end{equation}

\noindent where

\vspace{-3mm}

\begin{equation}
A_{2}^{r}\left(\phi_{n}\right)=2\Re\left\{ e^{-j\phi_{n}}\bar{d}_{n}^{r}\right\} ,\label{eq: A_2_phi}
\end{equation}

\noindent and

\vspace{-3mm}

\begin{equation}
A_{2}^{r}\left(\Phi^{r}\backslash\phi_{n}\right)=2\Re\left\{ \sum_{i\neq n}^{N}e^{-j\phi_{i}}\bar{d}_{i}^{r}\right\} .\label{eq: A_2_without_phi}
\end{equation}

Hence, $A_{1}^{r}\left(\Phi^{r}\backslash\phi_{n}\right)$ in (\ref{eq: A_1_without_phi}),
$A_{2}^{r}\left(\Phi^{r}\backslash\phi_{n}\right)$ in (\ref{eq: A_2_without_phi})
and $\left\Vert \boldsymbol{h}^{r}\right\Vert ^{2}$ in (\ref{eq: Rx SNR-LC-OA-1})
can be neglected in solving (\ref{eq: theta_AO}) since they are independent
of $\phi_{n}$. Consequently, using (\ref{eq: A_1_phi}) and (\ref{eq: A_2_phi}),
the solution of (\ref{eq: theta_AO}) yields

\vspace{-3mm}

\begin{equation}
\phi_{n}^{*}=\hspace{-1mm}\hspace{-1mm}\underset{\begin{array}{c}
\phi_{n,t}\in\mathcal{L}\end{array}}{\text{arg}\,\text{\,max}}\hspace{-1mm}\Re\left\{ \hspace{-1mm}e^{-j\phi_{n,t}}\underset{\text{Term 3}}{\underbrace{(\bar{d}_{n}^{r}+\sum_{k\neq n}^{N}e^{j\phi_{k,t-1}}(d_{k,n}^{r})^{*})}}\hspace{-1mm}\right\} \hspace{-1mm},\label{eq: theta_LC-AO}
\end{equation}

\noindent where $d_{k,n}^{r}$ and $\bar{d}_{n}^{r}$ are given by
(\ref{eq: d_k_n}) and (\ref{eq: d_k_hat}), respectively. Iterations
are also required to improve the performance of (\ref{eq: theta_LC-AO}).
It is worth noting that the LC-AO algorithm calculates Term 3 in (\ref{eq: theta_LC-AO})
only once for all $2^{b}$ possible combinations of $\phi_{n}$. Algorithm
\ref{alg:The-proposed- LC-AO} summarizes the proposed LC-AO algorithm.

\begin{algorithm}[h]
\begin{itemize}
\item \textbf{Input:} $\bar{\boldsymbol{g}}_{n}^{r}$, $\boldsymbol{G}^{r}$,
$\boldsymbol{h}^{r}$, $\mathcal{L}$, and maximum \# of iteration,
$T$;
\item \textbf{Initiate:}{\small{} $\phi_{n,0}=0$} for \textbf{$n=1,\ldots,N$};
\end{itemize}
~~~~~1: \textbf{Set: $t\leftarrow1$};

~~~~~2: \textbf{for$\,\,\,r=1$ to $R$}

~~~~~3:\textbf{ ~~~while} \textbf{$\,\,\,t\leq T$, do}

~~~~~4:\textbf{ ~~~~~~~~~for$\,\,\,n=1$ to $N$}

~~~~~5:\textbf{ ~~~~~~~~~~~~Buffer: $\zeta_{max}=[\cdot]$},
and \textbf{$\psi_{n}^{r}=[\cdot]$};

~~~~~6:\textbf{ ~~~~~~~~~~~~for$\,\,\,k=1$ to $N$}

~~~~~7:\textbf{ ~~~~~~~~~~~~~~~if$\,\,\,k\neq n$,
do}

~~~~~8:\textbf{ ~~~~~~~~~~~~~~~~~$d_{k,n}^{r}=\sum_{i=1}^{d_{f}}\left(g_{k,i}^{r}\bar{g}_{k}^{r}\right)\left(g_{n,i}^{r}\bar{g}_{n}^{r}\right)^{*}$};

~~~~~9:\textbf{ ~~~~~~~~~~~~~~~~~Set: $\psi_{n}^{r}\leftarrow\psi_{n}^{r}+e^{j\phi_{k,t-1}}(d_{k,n}^{r})^{*}$};

~~\hspace*{1mm}\hspace*{1mm}10:\textbf{ ~~~~~~~~~~~~~~~end
if}

~~\hspace*{1mm}\hspace*{1mm}11:\textbf{ ~~~~~~~~~~~~end
for}

~~\hspace*{1mm}\hspace*{1mm}12:\textbf{ ~~~~~~~~~~~~Compute:
$\bar{d}_{n}^{r}=\sum_{i=1}^{d_{f}}\left(g_{n,i}^{r}\bar{g}_{n}^{r}\right)\left(h_{i}^{r}\right)^{*}$};

~~\hspace*{1mm}\hspace*{1mm}13:\textbf{ ~~~~~~~~~~~~Compute:
$\text{Term 3}=\bar{d}_{n}^{r}+\psi_{n}^{r}$};

~~\hspace*{1mm}\hspace*{1mm}14:\textbf{ ~~~~~~~~~~~~for$\,\,\,l=1$
to $2^{b}$}

~~\hspace*{1mm}\hspace*{1mm}15:\textbf{ ~~~~~~~~~~~~~~~~$\zeta_{max}(1,l)\leftarrow\Re\left\{ e^{-j\phi_{n,t,l}}\times\text{Term 3}\right\} $};

~~\hspace*{1mm}\hspace*{1mm}16:\textbf{ ~~~~~~~~~~~~end
for}

~~\hspace*{1mm}\hspace*{1mm}17:\textbf{ ~~~~~~~~~~~~Find:
$i_{max}=\text{arg\,\,max}\{\zeta_{max}\}$};

~~\hspace*{1mm}\hspace*{1mm}18: \textbf{~~~~~~~~~~~~Update:
$\phi_{n,t}\leftarrow\mathcal{L}(i_{max})$};

~~\hspace*{1mm}\hspace*{1mm}19:\textbf{ ~~~~~~~~~end
for}

~~\hspace*{1mm}\hspace*{1mm}20:\textbf{ ~~~~~~~~~Set:
$t\leftarrow t+1$};

~~\hspace*{1mm}\hspace*{1mm}21: \textbf{~~~end while}

~~\hspace*{1mm}\hspace*{1mm}22: \textbf{~~~Set: $\Phi^{r}\leftarrow\text{diag}([e^{-j\phi_{1,T}}\ldots e^{-j\phi_{N,T}}])$}

~~\hspace*{1mm}\hspace*{1mm}23: \textbf{end for}
\begin{itemize}
\item \textbf{Output}{\small{} }$\Phi^{r}$ for $r=1,\ldots,R$.
\end{itemize}
\caption{\label{alg:The-proposed- LC-AO}The proposed LC-AO algorithm pseudo-code.}
\end{algorithm}

\section{\label{sec:Complexity-Analysis}Complexity Analysis}

The computational complexity of the proposed AO and LC-AO algorithms
is derived in terms of the real additions and real multiplications.
For the AO algorithm, the matrix and vector operation inside the norm
in (\ref{eq: theta_AO}) requires $2N(2d_{f}+1)$ and $4N(d_{f}+1)$
real additions and multiplications, respectively, for each possible
value of $\phi_{n}$. Furthermore, the second norm performs $(2d_{f}-1)$
and $2d_{f}$ real additions and multiplications, respectively. Thus,
the total number of real additions, $\text{RA}_{\text{AO}}$, and
multiplications, $\text{RM}_{\text{AO}}$, required to perform the
AO algorithm is respectively given as

\vspace{-3mm}

\begin{equation}
{\color{red}{\color{black}\text{RA}_{\text{AO}}}}=RN2^{b}\left(2N\left(2d_{f}+1\right)+2d_{f}-1\right),\label{eq: RA_OA}
\end{equation}

\vspace{-3mm}

\begin{equation}
{\color{red}{\color{black}\text{RM}_{\text{AO}}}}=RN2^{b}\left(4N\left(d_{f}+1\right)+2d_{f}\right).\label{eq: RM_OA}
\end{equation}

In the proposed LC-AO algorithm, (\ref{eq: d_k_n}) and (\ref{eq: d_k_hat})
require $(8d_{f}-2)$ and $(6d_{f}-2)$ real additions, as well as
$12d_{f}$ and $8d_{f}$ real multiplications, respectively. Thus,
Term 3 in (\ref{eq: theta_LC-AO}) requires $N(8d_{f}+1)-2(d_{f}+1)$
and $4N(3d_{f}+1)-4(d_{f}+1)$ real additions and multiplications,
respectively. It is worth noting that the LC-AO algorithm calculates
Term 3 only once for all $2^{b}$ possibilities of $\phi_{n}$. Therefore,
the total number of real additions, $\text{RA}_{\text{LC-AO}}$, and
multiplications, $\text{RM}_{\text{LC-AO}}$, required to perform
the LC-AO algorithm are respectively given as

\vspace{-3mm}

\begin{equation}
{\color{red}{\color{black}\text{RA}_{\text{LC-AO}}}}=RN\left(2^{b+1}+N\left(8d_{f}+1\right)-2\left(d_{f}+1\right)\right),\label{eq: RA_OA-1}
\end{equation}

\vspace{-3mm}

\begin{equation}
{\color{red}{\color{black}\text{RM}_{\text{LC-AO}}}}=RN\left(2^{b+2}+4N\left(3d_{f}+1\right)-4\left(d_{f}+1\right)\right).\label{eq: RM_OA-1}
\end{equation}

\vspace{-5mm}

\section{\label{sec:Simulation-Results}Simulation Results}

\begin{figure}
\noindent \begin{centering}
\includegraphics[scale=0.41]{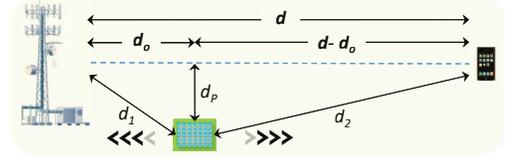}
\par\end{centering}
\begin{raggedright}
\caption{\label{fig:Simulation-setup.}Simulation setup.}
\par\end{raggedright}
\vspace{-3mm}
\end{figure}

In this section, Monte Carlo simulations are used to evaluate the
proposed AO and LC-AO algorithms for uplink RIS-SCMA. The codebooks
for all users are set based on the approach in \cite{codebook_design_2014}.
The parameters employed in simulations are $U=6$, $R=4$, $d_{f}=3$,
$b=3$, $T=3$, and $M=2$. As depicted in Fig. \ref{fig:Simulation-setup.},
the simulation setup is as follows: $d=40$ m, $d_{p}=1.5$ m, $d_{1}=\sqrt{d_{p}^{2}+d_{o}^{2}}$
m, and $d_{2}=\sqrt{d_{p}^{2}+(d-d_{o})^{2}}$ m. The overall system
path loss is $(\lambda^{4}d_{1}^{-2}d_{2}^{-2})/(256\pi^{2})$ \cite{pathloss},
where $\lambda$ denotes the signal wavelength for the $2.4$ GHz
operating frequency. All channels are Rician fading with a factor
of $1$. The average received SNR of the RIS-SCMA system for all $R$
OREs, $\Gamma$, in dB is used to assess the behavior of the proposed
optimization algorithms. The blind scenario mentioned in the simulation
results refers to the case where the RIS reflects the incident signals
without performing any optimization on the phase shifts (i.e., $\phi_{n}=0$,
$\forall n$ and $\forall r$). This scenario is included only to
show the gain obtained from comparing the optimized scenario with
the non-optimized one.

\begin{figure}
\begin{centering}
\includegraphics[scale=0.41]{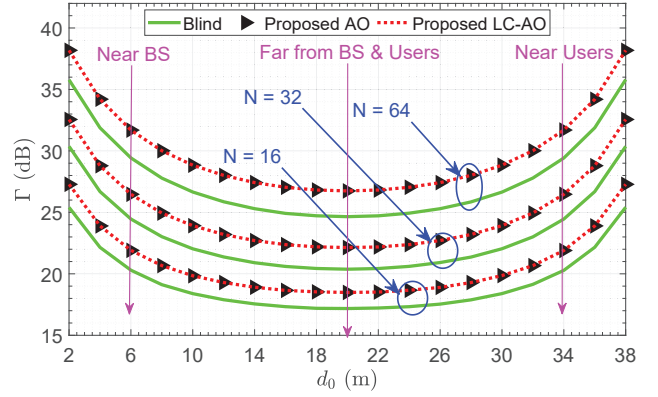}
\par\end{centering}
\begin{raggedright}
\caption{\label{fig:Rate-vs.-}Deployment investigation of the RIS for the
uplink RIS-SCMA system.}
\par\end{raggedright}
\vspace{-5mm}
\end{figure}

Fig. \ref{fig:Rate-vs.-} considers the deployment of the RIS for
the uplink RIS-SCMA system. This investigation involves varying $d_{0}$
(from the BS towards the users) for different values of the RIS elements,
i.e., $N=16$, 32 and 64. It is shown that the maximum average received
SNR occurs when the RIS is either near the BS or near the users. In
this letter, we consider that the RIS is placed near the BS with $d_{0}=2$
m. Further, results also show that the proposed AO and LC-AO algorithms
provide the same received SNR and outperform the blind scenario.

\begin{figure}
\begin{centering}
\includegraphics[scale=0.41]{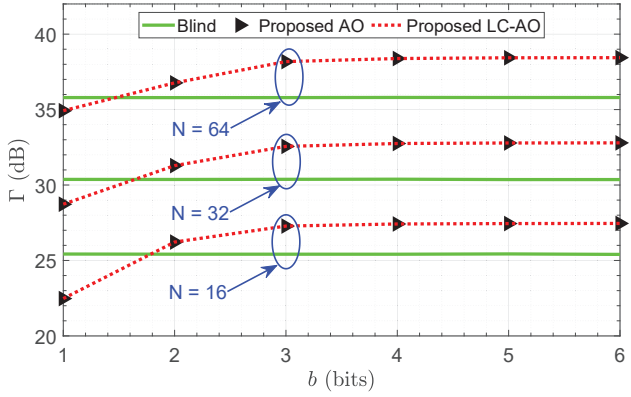}
\par\end{centering}
\vspace{-1mm}

\caption{\label{fig:Rate-versus-phases}Effect of the quantization bits for
each RIS phase shift.}

\vspace{-5mm}
\end{figure}

Fig. \ref{fig:Rate-versus-phases} depicts the average received SNR
at $d_{0}=2$ m versus different RIS phase shift values for different
values of $N$. The proposed AO and LC-AO algorithms provide the same
received SNR, and outperform the blind scenario in terms of the received
SNR when $b\geq2$. As seen from Fig. \ref{fig:Rate-versus-phases},
the improvement of the received SNR of the proposed algorithms saturates
when $b\geq3$. Furthermore, the computational complexity of the proposed
algorithms increases as $b$ increases, according to Section \ref{sec:Complexity-Analysis}.
Thus, $b=3$ is selected. It should be pointed out that for $b=1$,
only two phase shift values are available for each RIS element (i.e.,
$\mathcal{L}=\{-\pi,0\}$ in (\ref{eq: P1_C1})), which is not enough
to improve the performance over the blind scenario.

\begin{figure}
\begin{centering}
\hspace*{-3mm}\subfloat[\textcolor{red}{\label{fig:Convergence-speed.}}Convergence speed.]{\begin{centering}
\includegraphics[scale=0.305]{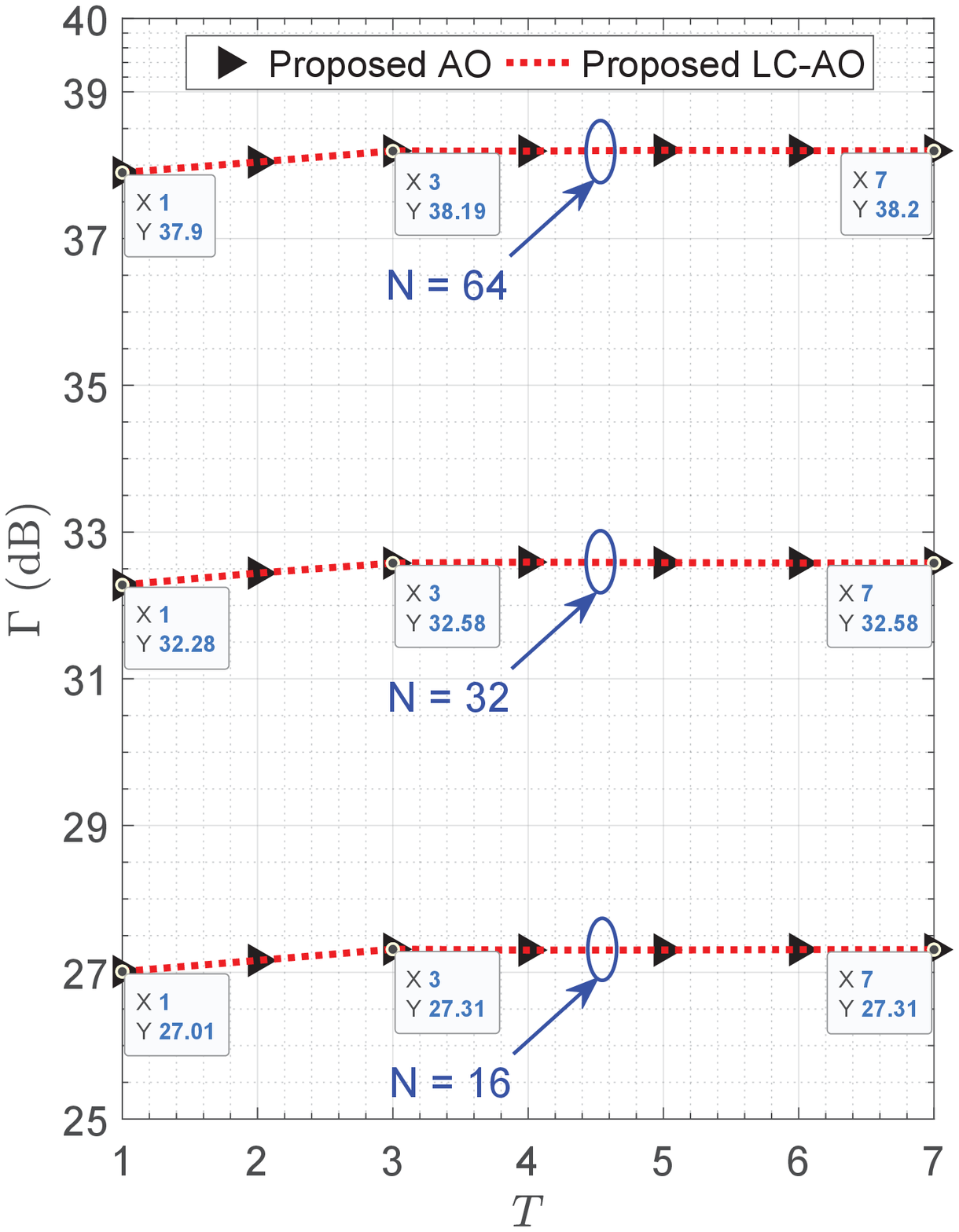}
\par\end{centering}
}\hspace*{-3mm}\subfloat[\label{fig:Effect-of}Effect of $N$.]{\begin{centering}
\includegraphics[scale=0.305]{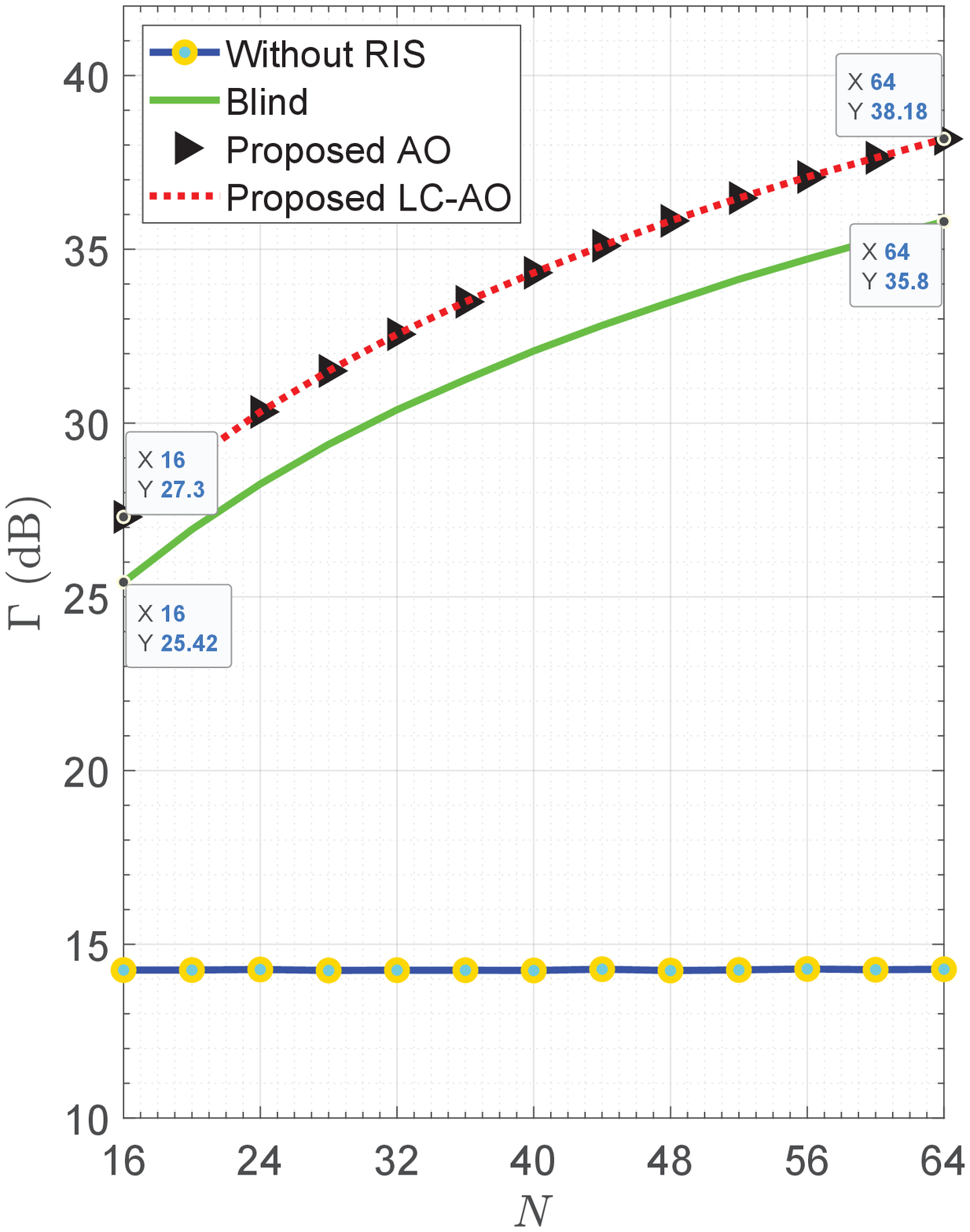}
\par\end{centering}

}
\par\end{centering}
\caption{\label{fig:Rate-versus- N} Convergence speed and effect of $N$ on
the proposed algorithms for the uplink RIS-SCMA system.}

\vspace{-5mm}
\end{figure}

Fig. \ref{fig:Convergence-speed.} shows that only three iterations
(i.e., $T=3$) are enough for the proposed algorithms to converge.
Therefore, at $d_{o}=2$ m, $b=3$ and $T=3$, the average received
SNR of the proposed AO and LC-AO algorithms is the same and increases
as $N$ increases, as shown in Fig. \ref{fig:Effect-of}. It is also
seen that the proposed algorithms provide a significant improvement
in the received SNR compared to the blind and without RIS scenarios;
e.g., the proposed algorithms provide $1.88$ dB and $2.38$ dB improvement
in the received SNR at $N=16$ and $64$, respectively, compared to
the blind scenario.

\begin{figure}
\begin{centering}
\subfloat[\label{fig:. complexity b}$b=3$.]{\begin{centering}
\includegraphics[scale=0.35]{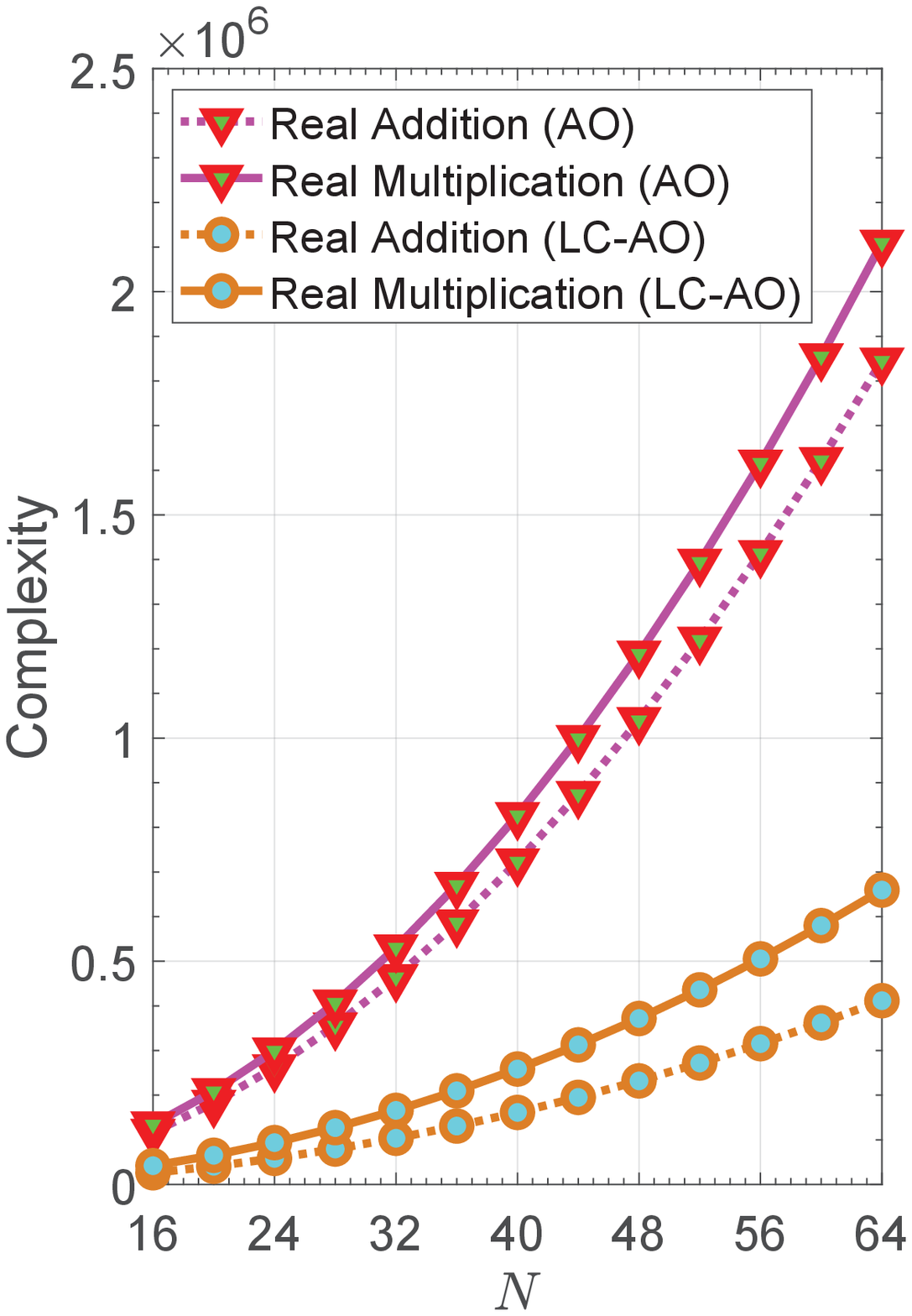}
\par\end{centering}
}\subfloat[\label{fig:. complexity N}$N=32$.]{\begin{centering}
\includegraphics[scale=0.35]{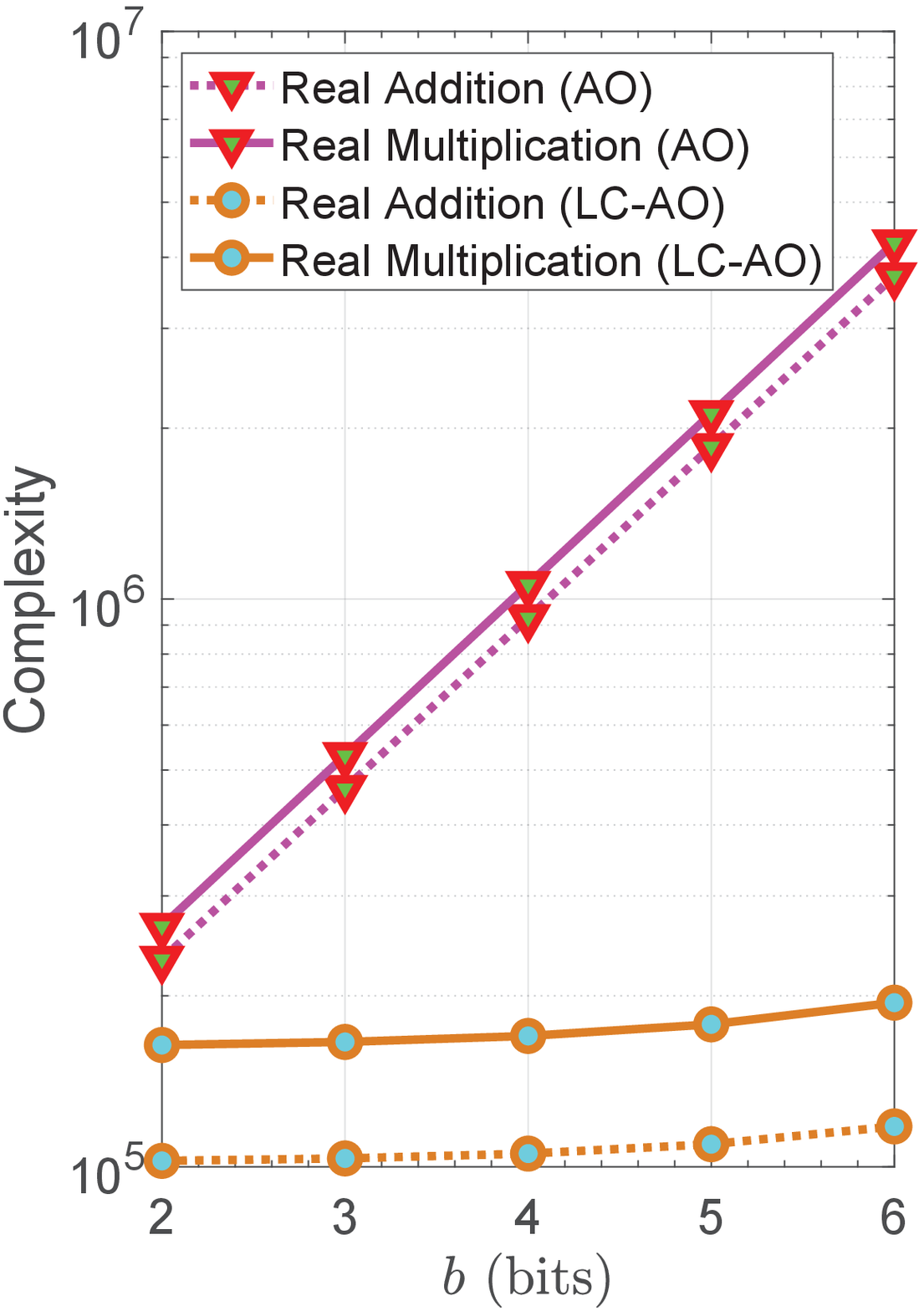}
\par\end{centering}
}
\par\end{centering}
\caption{\label{fig:Complexity-of-the}Complexity comparison of the proposed
algorithms.}

\vspace{-5mm}
\end{figure}

As seen from Fig. \ref{fig:Complexity-of-the} and Section \ref{sec:Complexity-Analysis},
the proposed LC-AO algorithm significant reduces the computational
complexity in terms of the number of real additions and multiplications.
The complexity of the AO algorithm significantly increases as $N$
increases, while that of the LC-AO algorithm increases less with $N$,
as illustrated in Fig. \ref{fig:. complexity b}. Furthermore, unlike
the AO algorithm, the complexity of the LC-AO algorithm slightly increases
as $b$ increases, as shown in Fig. \ref{fig:. complexity N}.

Finally, the proposed algorithms provide up to $2.4$ dB improvement
in the received SNR compared with the blind scenario. Besides, the
complexity of the proposed LC-AO algorithm is significantly lower
when compared with that of the AO algorithm for the same received
SNR.

\section{\label{sec:Conclusion}Conclusion}

For the first time, this letter has proposed an alternate optimization
algorithm to optimize the uplink RIS-SCMA discrete phase shifts, namely,
the AO algorithm. Furthermore, a novel variant of the AO algorithm
has been introduced, referred to as the LC-AO algorithm; this has
achieved the same received SNR as the AO algorithm, but with a significant
reduction in computational complexity. The deployment and optimal
number of discrete phase shifts for RIS have been investigated. The
simulation results have shown that placing the RIS near the BS or
users is beneficial, while the 3-bit quantization for RIS phase shifts
is enough to improve the received SNR with an acceptable complexity.
The proposed algorithms have significantly improved the received SNR
compared to the blind scenario, and this improvement increases with
the number of RIS elements. The system model can be extended to include
multiple antennas for the BS and users in the future work.

\end{document}